\documentclass[pra,superscriptaddress,twocolumn,longbibliography]{revtex4-1}

\usepackage{amsmath}
\usepackage{amssymb}
\usepackage{amsxtra,color}
\usepackage{latexsym}
\usepackage{graphicx}
\usepackage{tikz}
\usepackage{pgfplots}
\usepgfplotslibrary{groupplots}
\pgfplotsset{compat=1.3}
\usepackage{subcaption}
\usepackage{lipsum}
\usepackage{color}
\usepackage{braket}
\usepackage{lineno}
\usepackage{ulem}

\usepackage{wasysym}

\DeclareMathOperator*{\intsum}{%
\mathchoice%
  {\ooalign{$\displaystyle\sum$\cr\hidewidth$\displaystyle\int$\hidewidth\cr}}
  {\ooalign{\raisebox{.14\height}{\scalebox{.7}{$\textstyle\sum$}}\cr\hidewidth$\textstyle\int$\hidewidth\cr}}
  {\ooalign{\raisebox{.2\height}{\scalebox{.6}{$\scriptstyle\sum$}}\cr$\scriptstyle\int$\cr}}
  {\ooalign{\raisebox{.2\height}{\scalebox{.6}{$\scriptstyle\sum$}}\cr$\scriptstyle\int$\cr}}
}

\usepackage[utf8]{inputenc}
\usepackage{MnSymbol}	% use for sumint

\usepackage{hyperref}

\definecolor{MyDarkGreen}{rgb}{0,0.6,0}
\definecolor{MyDarkBlue}{rgb}{0,0,0.8}
\definecolor{MyDarkRed}{rgb}{0.6,0,0.3}

\begin{document}

\title{Resonant two-photon ionization of helium atoms studied by attosecond interferometry}

\author{L. \surname{Neori\v{c}i\'c}}
\thanks{These two authors contributed equally}
\email{lana.neoricic@fysik.lth.se}
\affiliation{Department of Physics, Lund University, Box 118, 22100 Lund, Sweden}

\author{D. \surname{Busto}}
\thanks{These two authors contributed equally}
\email{david.busto@fysik.lth.se}
\affiliation{Department of Physics, Lund University, Box 118, 22100 Lund, Sweden}
\affiliation{Institute of Physics, Albert Ludwig University, Stefan-Meier-Strasse 19, 79104  Freiburg, Germany}

\author{H. \surname{Laurell}}
\affiliation{Department of Physics, Lund University, Box 118, 22100 Lund, Sweden}

\author{R. \surname{Weissenbilder}}
\affiliation{Department of Physics, Lund University, Box 118, 22100 Lund, Sweden}

\author{M. Ammitzb{\"o}ll}
\affiliation{Department of Physics, Lund University, Box 118, 22100 Lund, Sweden}

\author{S.~\surname{Luo}}
\affiliation{Department of Physics, Lund University, Box 118, 22100 Lund, Sweden}

\author{J.~Peschel}
\affiliation{Department of Physics, Lund University, Box 118, 22100 Lund, Sweden}

\author{H. Wikmark}
\affiliation{Department of Physics, Lund University, Box 118, 22100 Lund, Sweden}

\author{J. Lahl}
\affiliation{Department of Physics, Lund University, Box 118, 22100 Lund, Sweden}

\author{S. Maclot}
\affiliation{Department of Physics, Lund University, Box 118, 22100 Lund, Sweden}

\author{R. J. \surname{Squibb}}
\affiliation{Department of Physics, University of Gothenburg, Origov\"agen 6B, 41296 Gothenburg, Sweden}

\author{S. \surname{Zhong}}
\affiliation{Department of Physics, Lund University, Box 118, 22100 Lund, Sweden}

\author{P.~\surname{Eng-Johnsson}}
\affiliation{Department of Physics, Lund University, Box 118, 22100 Lund, Sweden}

\author{C. L. \surname{Arnold}}
\affiliation{Department of Physics, Lund University, Box 118, 22100 Lund, Sweden}

\author{R. \surname{Feifel}}
\affiliation{Department of Physics, University of Gothenburg, Origov\"agen 6B, 41296 Gothenburg, Sweden}

\author{M. \surname{Gisselbrecht}}
\affiliation{Department of Physics, Lund University, Box 118, 22100 Lund, Sweden}

\author{E. Lindroth}
\affiliation{Department of Physics, Stockholm University, AlbaNova University Center, SE-106 91 Stockholm, Sweden}

\author{A. \surname{L'Huillier}}
\affiliation{Department of Physics, Lund University, Box 118, 22100 Lund, Sweden}
\email{anne.lhuillier@fysik.lth.se}


\begin{abstract}
We study resonant two-photon ionization of helium atoms via the $1s3p$, $1s4p$ and $1s5p^1$P$_1$ states using the 15$^\mathrm{th}$ harmonic of a titanium-sapphire laser for the excitation and a weak fraction of the laser field for the ionization. The phase of the photoelectron wavepackets is measured by an attosecond interferometric technique, using the 17$^\mathrm{th}$ harmonic. We perform experiments with angular resolution using a velocity map imaging spectrometer and with high energy resolution using a magnetic bottle electron spectrometer. Our results are compared to calculations using the two-photon random phase approximation with exchange to account for electron correlation effects. We give an interpretation for the multiple $\pi$-rad phase jumps observed, both at and away from resonance, as well as their dependence on the emission angle.  
\end{abstract}

\maketitle

\section{Introduction}

Attosecond techniques using single pulses \cite{Goulielmakis,Schultze} or pulse trains \cite{Paul,Klunder} combined with a synchronized laser field have become essential tools for the study of photoionization processes in atoms and molecules. The phase information provided by techniques such as streaking  \cite{Kienberger} or RABBIT (Reconstruction of Attosecond Beating By Interference of Two-photon transitions)  \cite{Paul} complements cross-section measurements, giving insight into the photoionization temporal dynamics \cite{Schultze,Klunder}. Experiments have been performed in different atomic \cite{Ossiander,Zhong2020,AlexandridiPRR2021} and molecular \cite{haessler,Worner,Nandi2020} systems, and in broad photon energy regions, typically between $20$ and $100$ eV \cite{Isinger,Ossiander}. When atoms or molecules are ionized into ``flat'', structureless continua, the phase does not vary much with energy, apart from the threshold region, which may be affected by the variation of the so-called Coulomb phase and a laser-induced ``continuum-continuum'' phase \cite{Pazourak,dahlstrom}. In contrast, pronounced phase variations are observed in the vicinity of Cooper minima \cite{Klunder,AlexandridiPRR2021} and broad shape resonances in atomic or molecular systems \cite{Zhong2020,Nandi2020}.

These studies require broadband radiation, single attosecond pulses or attosecond pulse trains, typically spanning a few tens of eV in energy, in order to resolve ultrafast dynamics with time scales in the attosecond regime. Attosecond techniques may also be applied to the study of two-photon resonant processes, where the first, extreme ultraviolet (XUV) photon comes into resonance with a bound  \cite{Swoboda,Villeneuve2017,Drescher2022,autuori} or quasi-bound autoionizing state~\cite{haessler,Barreau_2019}.
The study of resonant two-photon ionization usually demands a higher spectral resolution than that of non-resonant processes, and the time scales are in the femtosecond range. In the early measurements \cite{Swoboda,Kotur}, the XUV frequency was tuned the across the resonance. Recently, powerful energy-resolved methods like ``rainbow'' RABBIT have been developed \cite{gruson}. In particular, the nontrivial phase variation around the $2s2p$ doubly excited state in helium \cite{gruson,busto2018}, and the $3s^{-1}4p$ window resonance in argon \cite{Turconi} have been extensively studied. In simple cases like helium, the temporal dynamics of the wavepacket created by resonant photoionization can be recovered \cite{gruson,busto2018}.

Angle detection brings a new dimension, momentum, to this type of measurement, allowing the reconstruction of coherent electron wavepackets in time and space. This has been beautifully shown for two-photon resonant ionization of Ne via the $3p^53d^1\text{P}_1$
state \cite{Villeneuve2017}, and recently of He via the $1s3p$ and $1s4p^1\text{P}_1$ states \cite{autuori}. Photoionization studies with angular resolution \cite{Heuser,Cirelli2018,busto,autuori} 
have pointed out that additional phase jumps as a function of emission angle appear due to the interference of angular channels, beside the phase jumps as a function of energy. 

Here, we study two-photon resonant ionization of helium through intermediate states in the 1s\textit{n}p $^1\text{P}_1$ Rydberg series with $n=3-5$. We generate odd high-order harmonics of an infrared (IR) laser field, and tune the IR laser frequency in order to reach the $1s3p$, $1s4p$ or a coherent superposition of $1s4p$ and $1s5p$ states with the 15$^\mathrm{th}$ harmonic. We further ionize by absorption of an additional IR photon. We analyze the phase of the created wavepacket by interferometry using another, non-resonant two-photon process leading to the same final state, namely, absorption of the 17$^\mathrm{th}$ harmonic and emission of an IR photon. We perform two series of experiments, one with angular resolution and moderate energy resolution using a velocity map imaging spectrometer (VMIS) and one with high energy resolution and no angular resolution using a magnetic bottle electron spectrometer (MBES). In the latter case, we also rotate the polarization of the probe field in order to eliminate one of the angular channels. We discuss the theory of ``below-threshold'' interferometry in the angle-integrated and angle-resolved cases and we compare our experimental results with simulations using the two-photon random phase approximation with exchange (RPAE)~\cite{vinbladh2019} to account for electron correlation effects. This allows us to interpret the multiple phase jumps observed both as a function of energy and emission angle. 

\section{Principle of the experiment}

\begin{figure}
    \centering
   \includegraphics[width=\columnwidth]{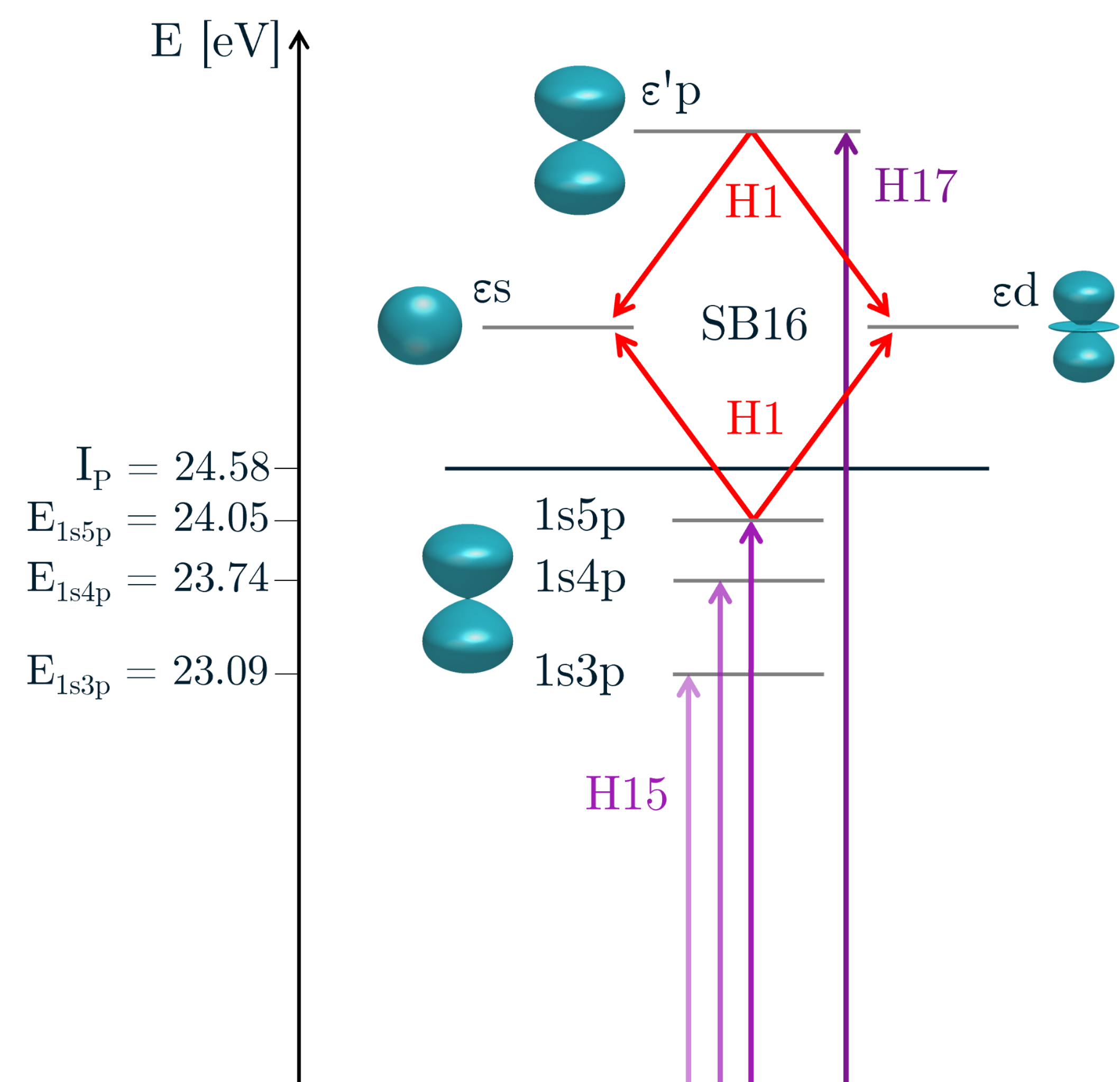}
    \caption{Energy levels and excitation processes involved in the present study. The violet (red) arrows denote excitation by the $15^{\text{th}}$ and $17^{\text{th}}$ harmonics (fundamental). The spherical harmonics corresponding to the different angular momentum channels are indicated in cyan.}
    \label{helium}
\end{figure}

Our experiment is based on the RABBIT technique, with a focus on a narrow spectral region just below the ionization threshold ($I_p$) of helium, as illustrated in Fig.\;\ref{helium}. The $15^{\text{th}}$ harmonic (H15) of the fundamental laser field is tuned into resonance with one or several Rydberg states. Harmonics of order 17 and above have enough energy to ionize the system and generate a comb of photoelectron peaks spaced by $2\hbar \omega$. In the presence of a dressing field with photon energy $\hbar\omega$, two quantum paths  lead to the same final state in the continuum, resulting in the formation of sidebands (SBs). 
Varying the relative time-delay between the XUV and IR pulses causes periodic modulations in the SB16 signal, which can generally be written as
\begin{equation}
    I_\mathrm{SB} = |A^+|^2 +|A^-|^2 + 2|A^+||A^- |\cos({2\omega\tau-\Delta\varphi)},
    \label{SB16}
\end{equation}
where $A^\pm$ denotes the amplitude of the absorption ($+$) and emission ($-$) paths. 
The oscillation phase can be decomposed as $\Delta\varphi=\Delta\varphi_\text{XUV}+\Delta\varphi_\text{A}$, where $\Delta\varphi_\text{XUV}$ is the group delay of the attosecond pulses and 
$\Delta\varphi_\text{A}$ originates from the two-photon ionization process. SB16 differs from the other sidebands as the absorption path is below the ionization threshold. In a first set of experiments, we record the SB16 signal as a function of energy and angle. In a second one, we perform an angle-integrated measurement with high spectral resolution. 
In both cases, we extract both the amplitude of the signal oscillating at frequency $2\omega$ and the phase of the oscillation by applying a spectrally resolved RABBIT analysis \cite{Paul,MullerOL,gruson}. 
In this implementation of the RABBIT technique, sometimes called ``below-threshold RABBIT'' \cite{Swoboda,kheifets,Drescher2022,autuori}, the interesting information is the phase change due to resonant two-photon ionization, which is ``read'' by interferometry with the emission path (absorption of the 17$^\mathrm{th}$ harmonic and emission of an IR photon).  

\section{Experimental method}

\begin{figure}
    \centering
   \includegraphics[width=\linewidth]{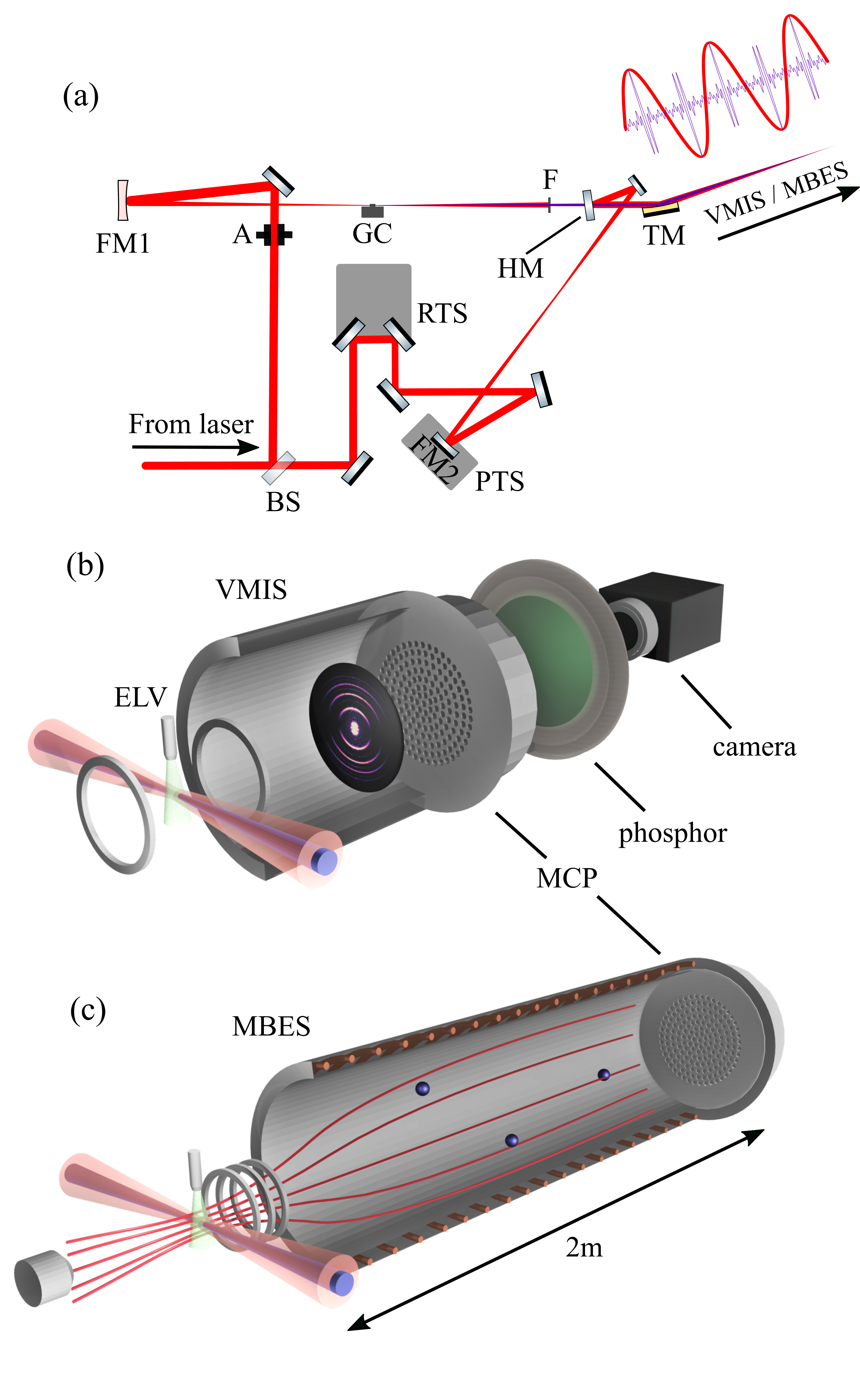}
    \caption{Experimental setup: (a) Optical interferometer allowing us to create an attosecond pulse train and a delayed dressing field; (b) Velocity map imaging spectrometer; (c) Magnetic bottle electron spectrometer. We use the following  abbreviations: BS, beam splitter; A, aperture; FM, focusing mirror; GC, gas cell; RTS, rough translation stage; PTS, piezo-driven translation stage; HM, holey mirror; F, filter; TM, toroidal mirror; ELV, Even-Lavie valve; MCP, microchannel plates.}
    \label{setup}
\end{figure}

A Ti:Sa-based laser operating at 1\,kHz provided 3\,mJ pulses with a central wavelength tunable from 780 to 820\,nm and a FWHM bandwidth of approximately 40\,nm. The pulses were split using a 70-30 beamsplitter, so that the intense fraction could be used as the pump beam and the weaker one as the probe, as illustrated in Figure \ref{setup}(a). The pump was focused into an 8 mm long pulsed gas cell containing a rare gas, thereby generating odd harmonics of the fundamental IR field. The resulting spectrum was filtered with a 200 nm thick aluminium filter so as to suppress the transmission of the pump beam into the detection region. The probe pulses were directed onto a piezomotor-driven delay stage and recombined with the XUV pulses in a helium jet. The intensity of the probe in the interaction region was reduced in order to avoid two-IR-photon absorption. 

Two complementary experimental measurements were performed, as indicated in Figure \ref{setup}(b,c). In the angle-resolved experiment (b), a velocity map imaging spectrometer \cite{VMI1997,RadingAS2018} was installed for the electron detection and the helium atoms were provided by an Even-Lavie valve pulsed at 500 Hz \cite{even}. To optimize the 15$^\mathrm{th}$ and 17$^\mathrm{th}$  harmonic signal, the gas target was chosen to be xenon or krypton. The photoelectrons were detected with a microchannel plate leading to a phosphor screen, recorded by a camera. The resulting images captured the projection of the momentum of electrons in the plane perpendicular to their travel. The momentum maps were recorded as a function of the delay between the pump and probe pulses, in steps of 250\,as. The photoelectron momentum distribution was obtained at each delay by performing an inverse Abel transform on the VMIS images \cite{inversion}. In order to minimize the appearance of mathematical artefacts in the centre of the inverted images, a well-known problem of inversion methods based on Legendre polynomial decomposition, an iterative inversion algorithm based on \cite{VrakkingRSI2001} was applied. The oscillation phase for each energy and angle bin was extracted by fitting a cosine function to the recorded time-delay signal.\\

The high spectral resolution measurements were carried out using an MBES for the electron detection, with a 2 m long flight tube and a 4$\pi$ sr collection angle [see Figure \ref{setup}(c)]. The energy resolution was better than 80 meV at low kinetic energies. He atoms were provided by an effusive gas jet. The delay between the two beams was stabilized with an RMS error of less than 20 as. High-order harmonics were generated in argon. In this experiment, the laser central wavelength was chosen to be 782 nm and the bandwidth 30 nm, so that the spectral region covered by the 15$^\mathrm{th}$  harmonic included the $1s4p^1$P$_1$ resonance. The relative contributions of \textit{s} and \textit{d} angular channels to the two-photon transition amplitude were varied by rotating the polarization of the probe in a direction either parallel or perpendicular to the polarization of the pump pulse.

\section{Theoretical method}

\subsection{Two-photon matrix element}
The two-photon transition matrix element connecting the initial state $\ket{a}$ with the  continuum state $\ket{q}$  via all dipole-allowed intermediate states can be written in the lowest order of perturbation theory as 
\begin{align}
\label{M2}
M^\pm_{qa}(\Omega,\omega)=
-i\lim_{\xi \rightarrow 0^+}
\intsum_p \frac{
\langle q \mid e z \mid p  \rangle \langle p  \mid    e z  \mid a \rangle}{\epsilon_a + \hbar \Omega -\epsilon_p + i \xi}  E_{q}E^\pm_{1}.
\end{align}
Here $\epsilon_a$ and $\epsilon_p$ represents the energies of the initial and intermediate states, $E_q$, $E_1^+=E_1$ the $q^\mathrm{th}$ harmonic and laser fields, with frequencies $\Omega$ and $\omega$, respectively. In this expression, electron correlation is neglected and lowest-order perturbation theory for the interaction with the radiation fields is assumed. In addition, both fields are described as monochromatic and linearly polarized along the same direction ($z$). Energy conservation implies that $\epsilon_q = \epsilon_a + \hbar\Omega \pm \hbar \omega$, where the sign $\pm$ refers to absorption or emission of the IR photon. Note that in the latter case, $E^-_1$ is the complex conjugate of $E_{1}^+$. Since the IR laser field is delayed by $\tau$ relative to the XUV field, $E_1^\pm \propto e^{\mp i\omega \tau}$. When the second photon absorption or emission is above the threshold ($\hbar \Omega > I_p$), the two-photon matrix element is intrinsically complex. When $\hbar \Omega < I_p$, which is the case when the 15$^\mathrm{th}$ harmonic is absorbed, the lowest-order contribution can be chosen to be real, but if electron correlation is accounted for, as discussed in section \ref{ecorr}, the matrix element is again intrinsically complex.

\subsection{Angular momentum channels}

The next step consists in using spherical coordinates and introducing the different angular momentum channels (see Figure \ref{helium}). Two-photon ionization of He leads to $s$- or $d$- final states, and the angle-resolved two-photon amplitude can be written as
\begin{align}
A_\parallel^\pm(\theta)  \propto \sum_{L=0,2}
%\nonumber \\ 
% \sum_{L_q}
  Y_{L, 0 } \left(\theta,0\right) 
\left( 
\begin{array}{ccc}
L & 1 & 1 \\
0 & 0  & 0
\end{array}
\right)
e^{i\left(\eta_{ L}-\tfrac{L\pi}{2}\right)} \mathbb{M}_{L}^\pm  
\label{A2}
\end{align}
Here $L={0,2}$ is the angular momentum of the final state, $\eta_{L}$ the energy-dependent scattering phase, sum of the Coulomb phase and a contribution from the short-range potential. The phase $L \pi/2$ describes the effect of the centrifugal potential. $Y_{L,0}$ denotes a spherical harmonic, and $\mathbb{M}_{L}^\pm$ is the reduced two-photon matrix element~\cite{Saha:2021}, which depends on the photon energies. 
The amplitude varies with the polar angle $\theta$, but due to the cylindrical symmetry, not with the azimuth. Since the ionization is from an $s$-orbital, the photoelectron can only occupy an $m=0$ state with respect to  the XUV-field polarization.

If the polarization of the laser field is turned 90$^{\circ}$ with respect to the XUV field, the interaction with the IR field will change the magnetic quantum number, as defined with respect to the XUV field,  by $\Delta m=\pm 1$. Since the ionization is from an $s$-orbital, only one angular momentum channel, the $d$-channel, survives.  
\begin{align}
A_\perp^\pm(\theta,\phi)  \propto 
 -Y_{2,1} \left(\theta,\phi \right)
\left( 
\begin{array}{ccc}
2 & 1 & 1 \\
-1 & 1  & 0
\end{array}
\right)
e^{i\eta_{2}} \mathbb{M}_2^\pm  
\label{A2p}
\end{align}
In this case, the angular dependence  will be that of a $d$-wave.
%, and the corresponding $\beta_i$ parameters will be constant, independent of energy or delay.

\subsection{Electron correlation}

 The two-color two-photon RPAE approach~\cite{vinbladh2019} is used to calculate correlated two-photon matrix elements that include channel coupling, after both one- and two-photon interaction, as well as ground state correlation. The method, which is described in detail in Ref.~\cite{vinbladh2019}, is gauge independent and was originally developed for above threshold ionization, but is here used in a scenario where, in the absorption path, no individual photon is able to ionize the atom by itself.  
To facilitate the comparison with experimental results, experimental energies are used for the bound excited states that are encountered in the sum over intermediate states $p$: $1s3p ^1\text{P}_1$, $1s4p ^1\text{P}_1$, and $1s5p ^1\text{P}_1$. Both time orders of the photon interaction are considered, but the result is completely dominated by the time order where the XUV photon is absorbed in the first step, as described by Eq.(\ref{M2}). The emission path of the RABBIT scan is over a smooth spectral region in the continuum and the phase change over the small energy interval discussed here is small. In the absorption path, the situation is very different. The bound state resonances are  narrow, with only light-induced broadening possible (see below).
If monochromatic XUV light is tuned over such an excited state resonance, the denominator in Eq.~(\ref{M2}) abruptly changes sign and a sharp $\pi$-rad jump of the phase is expected  at the resonance energy. Phase jumps are also to be expected when (and if) the matrix elements in the numerator change sign. 

\begin{figure}
    \centering
   \includegraphics[width=\columnwidth]{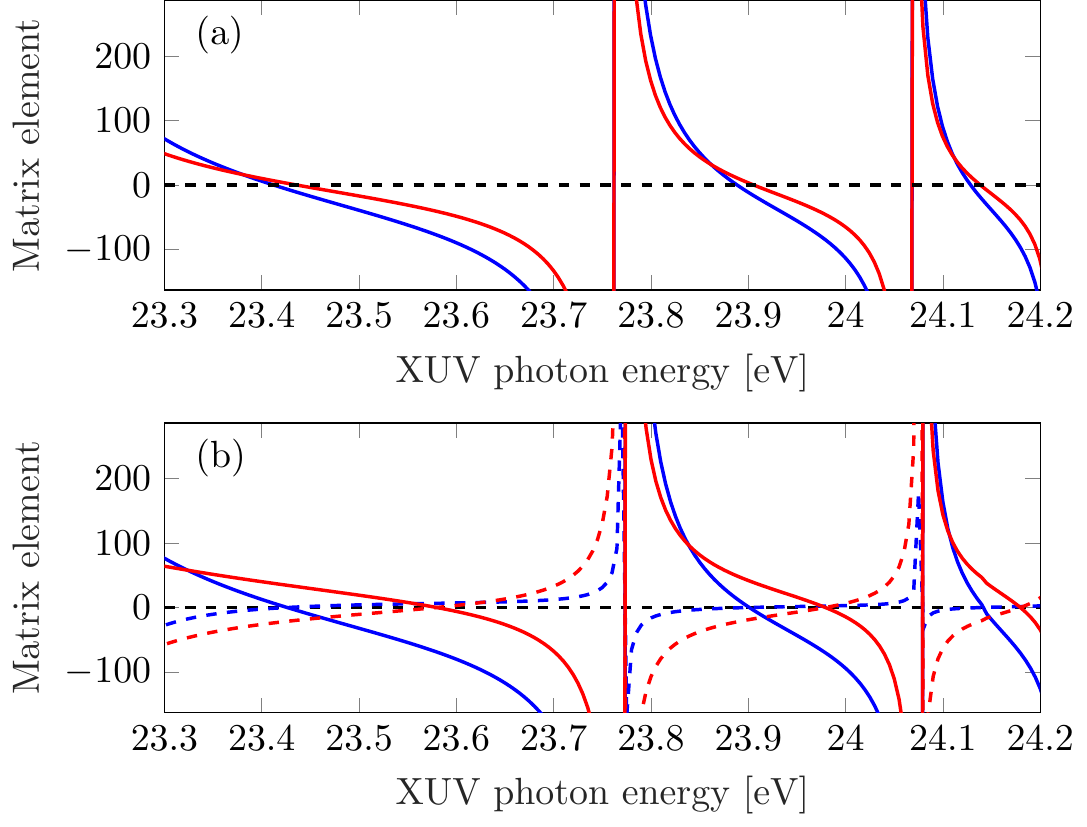}
    \caption{Absorption matrix elements to the $s$ (red) and $d$ (blue) continuum. The real and imaginary parts are shown respectively in the solid and dashed lines. The matrix elements in (a) do not include electron correlation while the matrix elements in (b) are obtained using RPAE.}
    \label{ME}
\end{figure}

Figure~\ref{ME} presents the real and imaginary parts of the two-photon transition amplitudes in the absorption path for the $s$ and the $d$ channels without (a) and with (b) electron correlation. The transition matrix elements from the ground state to the bound excited states and from these states to the continuum can be chosen to be real. Therefore, the amplitudes are real-valued in (a). At resonance, where the amplitude diverges, the phase variation is simply due to a sign change, while in between resonances it happens because the amplitude goes to zero. The phase jump in between the resonances happens at more or less the same energy for both the $s$ and the $d$ channel.

Through the introduction of electron correlation into the ionization process, the matrix elements acquire an imaginary part leading to a phase shift of the outgoing photoelectron. The phase variation at resonance is slightly smoother, while the energy where the phase jumps takes place between two resonances is now different for the $s$ and the $d$ channels. When incoherently adding the contributions from the two channels, this leads to a smoother phase variation between resonances. On the other hand, when these contributions are added coherently, the resulting phase jump depends on the angle of emission, as discussed more in detail below.  

\label{ecorr}

\subsection{Ionization-induced broadening and finite pulse effects}

It is possible to go beyond lowest-order perturbation theory for the dressing field close to resonance in a simple way, by adding an intensity-dependent AC-Stark shift ($\delta_p$) and ionization-induced width ($\gamma_p$) to the resonance energy ($\epsilon_p \rightarrow \epsilon_p+\delta_p+i\gamma_p$). The added complex term in the denominator will induce a change in the phase dependence across the resonance. Changing $\epsilon_p$ into $\epsilon_p+\delta_p+i\gamma_p$ into the denominator of Eq.~(\ref{M2}) leads to an additional phase term,  
\begin{equation}
    \chi_p=\mathrm{arctg}\left[\frac{\gamma_p}{\epsilon_a + \hbar \Omega -\epsilon_p-\delta_p}\right]. 
\end{equation}
Close to the resonance, the phase varies as $\mathrm{arctg}(\gamma_p/\delta \epsilon)$, where $\delta \epsilon$ goes through zero. The derivative of this function close to zero (and hence the slope of the phase variation with energy) is equal to $-1/\gamma_p$. Since $\gamma_p$ is proportional to the probe intensity, the absolute value of the slope of the phase variation across the resonance is expected to decrease as the intensity increases. In some of the simulations presented below, we include an ionization-induced width equal to 10 meV, corresponding to a probe intensity of   $6 \times 10^{11}$ W/cm$^{2}$. We do not include any Stark shift, however, since the origin of the energy axis in the experiment is not precisely known. Our experimental data are adjusted to the unshifted position of the resonant states. 

To compare with experimental results, we also include bandwidth effects through convolution with appropriate line profiles. Using Gaussian profiles defined by the central frequencies $\Omega_0$, $\omega_0$ and bandwidths $\sigma_q$, $\sigma_1$ for the XUV and IT fields respectively, the matrix elements leading to the same final energy $\epsilon_q$ are summed up as 
\begin{align}
\! M^{\pm}_{qa} \!\!= \!\!
\tfrac{1}{2 \pi \sigma_1 \sigma_q}\!\varint \!M^{\pm}_{qa}(\omega_{qa} \!\!\mp \! \omega, \omega) 
e^{-\frac{(\omega - \omega_0 )^2}{2 \sigma^2_1}-\frac{(\omega_{qa} \mp  \omega - \Omega_0 )^2}{2 \sigma^2_q}} \!d\omega,
\label{foldabs}
\end{align}
where $\hbar\omega_{qa}=\epsilon_q-\epsilon_a$. 

\section{Theory of below-threshold RABBIT in helium}

We now generalize the theory of RABBIT to the particular case where absorption of the lowest harmonic takes place in the discrete spectrum \cite{kheifets}. We consider two-photon ionization of He, with only a single angular channel for the first XUV photon absorption. 
In the angle-resolved case, the sideband signal is
\begin{equation}
    I_\text{SB}(\theta,\tau)= |A_\parallel^+(\theta,\tau)+A_\parallel^-(\theta,\tau)|^2, 
    \label{isb}
\end{equation}
where $\tau$ denotes the delay between the XUV and the IR fields, and $\theta$ is the emission angle. The amplitudes for the absorption and emission paths are defined as  
    \begin{equation}
        \!A_\parallel^\pm(\theta,\tau)  \!\propto\! -Y_{00}\sqrt{\tfrac{1}{3}}e^{i\eta_0}\mathbb{M}_0^\pm\!+\!Y_{20}(\theta,0)\sqrt{\tfrac{2}{15}}e^{i(\eta_2-\pi)}\mathbb{M}_2^\pm.
        \label{ang}
    \end{equation}
The coherent addition of the four terms included in Eqs.~(\ref{isb},\ref{ang}) depends on the phases  $\text{arg}[e^{i(\eta_L-L\pi/2)}\mathbb{M}_L^\pm]$. For the emission path, as in the ordinary RABBIT description with the intermediate state of the two-photon transition in the continuum, in the asymptotic limit, we have \cite{dahlstrom},
\begin{equation}
    \text{arg}\left(e^{i\left[\eta_L-\tfrac{L\pi}{2}\right]}\mathbb{M}_L^-\right)=\eta_1^-+\pi +\phi_{cc}^{L -}+\phi_\text{XUV}^--\omega \tau.
    \label{phase-em}
\end{equation}
Here $\eta_1^-$ denotes the scattering phase in the $p$-continuum, at the energy corresponding to the absorption of the 17$^\text{th}$ harmonic, $\phi_{cc}^{L -}$ is the additional phase due to the continuum-continuum emission process, which slightly depends on the angular momentum \cite{Fuchs:20,Peschel}, $\phi_\text{XUV}^-$ is the phase of the 17$^\text{th}$ harmonic and $\tau$ is the delay between the XUV and IR fields.  
For the absorption path,
\begin{equation}
    \text{arg}\left(e^{i\left[\eta_L-\tfrac{L\pi}{2}\right]}\mathbb{M}_L^+\right)=\eta_L-\tfrac{L\pi}{2}+\phi_L+\phi_\text{XUV}^++\omega \tau.
    \label{phase-abs}
    \end{equation}
$\eta_L$ is here the scattering phase of the final state ($s$ or $d$), $\phi_L$ reflects the sign changes of the two-photon matrix element at or in between resonances, as well as electron correlation (see Figure \ref{ME}) and $\phi_\text{XUV}^+$ is the phase of the 15$^\text{th}$ harmonic. 
In contrast to the case where both paths are above the threshold, the asymmetry between the two paths does not allow us to eliminate the scattering phase of the final state and extract the Wigner time delay \cite{dahlstrom}. A RABBIT measurement allows, however, the study of the phase variation ($\phi_L$) of the two-photon matrix element across the resonance.

\subsection{Angle-integrated below-threshold RABBIT}

\begin{figure}
    \centering
   \includegraphics[width=\columnwidth]{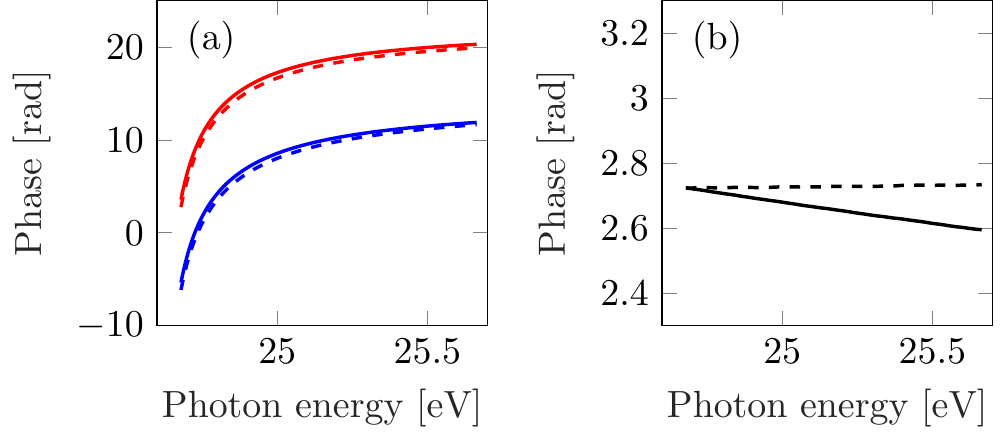}
    \caption{(a) Phase $\eta_L-L\pi/2$ (blue) and $\phi_\text{cc}^{L-}$ (red) as function of energy for $L=0$ (solid) and $L=2$ (dashed); (b) Reference phase $\varphi_\text{ref}^L=\eta_L-\eta_1^--\phi_\text{cc}^{L-}-L\pi/2-\pi$ for $L=0$ (solid) and $L=2$ (dashed).}
    \label{eta}
\end{figure}
In an angle-integrated measurement, the $\theta$-integration of $I_\text{SB}(\theta,\tau)$ eliminates the cross-products between different angular momentum channels (due to the orthogonality of the spherical harmonics) which now add incoherently,
\begin{align}
    I_\mathrm{SB}(\tau) &\propto \frac{1}{3}\left[|\mathbb{M}_0^+|^2 +|\mathbb{M}_0^-|^2 + 2|\mathbb{M}_0^+||\mathbb{M}_0^-|\cos(\Delta\varphi_0)\right]\nonumber\\
    &+ \frac{2}{15}\left[|\mathbb{M}_2^+|^2 +|\mathbb{M}_2^-|^2 + 2|\mathbb{M}_2^+||\mathbb{M}_2^-|\cos(\Delta\varphi_2)\right],
    \label{angle-int}
\end{align}
where 
\begin{align}
    \Delta \varphi_L= \text{arg}\left(e^{i\left[\eta_L-\tfrac{L\pi}{2}\right]}\mathbb{M}_L^+\right)-\text{arg}\left(e^{i\left[\eta_L-\tfrac{L\pi}{2}\right]}\mathbb{M}_L^-\right). 
\end{align}
In ordinary (above-threshold) RABBIT,  
\begin{equation}
    \Delta \varphi_L=2\omega\tau-\Delta \phi_\text{XUV}-\Delta \eta_1- \Delta \phi_\text{cc}^L,
\end{equation}
where $\Delta \eta_1= \eta_1^--\eta_1^+$, $\Delta \phi_\text{XUV}= \phi_\text{XUV}^--\phi_\text{XUV}^+$ and $\Delta \phi_\text{cc}^L= \phi_\text{cc}^{L-}-\phi_\text{cc}^{L+}$. Neglecting the small $L$-dependence of $\Delta \phi_\text{cc}$ \cite{Peschel}, the two oscillatory terms in Eq.~(\ref{angle-int}) are in phase, meaning that RABBIT measurement allows to unambiguously determine the phase $\Delta \eta_1$ \cite{Klunder,dahlstrom} and the Wigner delay approximated by $\Delta \eta_1/2\omega$. When one path is below threshold, however, 
\begin{equation}
    \Delta \varphi_L=2\omega\tau-\Delta \phi_\text{XUV}+\varphi_\text{ref}^L+\phi_L,
    \label{ref}
\end{equation}
with $\varphi_\text{ref}^L=\eta_L-\eta_1^--\phi_\text{cc}^{L-}-L\pi/2-\pi$.
Eq.~\ref{angle-int} describes the incoherent addition of two terms oscillating with the same frequency $2\omega$, but with different phase variation as a function of energy ($\phi_L$) and different reference phase ($\varphi_\text{ref}^L$), complicating the interpretation of the extracted phase. Although both $\eta_L$ and $\phi_\text{cc}^{L-}$ vary significantly close to the threshold due to the influence of the Coulomb potential, as shown in Figure~\ref{eta} (a), the reference phase $\varphi_\text{ref}^L$ does not depend much on the energy or on $L$ in the range investigated in the present work [Fig.~\ref{eta} (b)]. This implies that when $\phi_0\simeq \phi_2$, for example at resonance, both terms in Eq.~(\ref{angle-int}) vary with a similar phase offset. However, when $\phi_L$ changes sign because the $L$-dependent matrix element goes to zero, which happens at different energies for the $s$ and $d$ channels, the interpretation of the phase extracted from the experimental measurement is not clear. This is where angle-resolved measurements or the use of different polarizations for the excitation fields become useful.       

\subsection{Angle-resolved below-threshold RABBIT}

We first rewrite Eq.~(\ref{ang}) as
    \begin{equation}
        A_\parallel^\pm(\theta)  \propto e^{i\eta_0}\mathbb{M}_0^\pm+\sqrt{2}P_2(\cos\theta)e^{i\eta_2}\mathbb{M}_2^\pm,
        \label{ang2}
    \end{equation}
where we have introduced $P_2$, the Legendre polynomial of order 2, equal to $(3x^2-1)/2$.
Using Eqs.~(\ref{phase-em},\ref{phase-abs}), we have
 \begin{align}
       &A_\parallel^-(\theta) +A_\parallel^+(\theta)  \propto \label{bra} \\\nonumber
        &\ \ e^{i(-\omega\tau+\phi_{\text{XUV}}^-+\eta_1^-)}  \left[e^{i\phi_{\text{cc}}^{0-}}|\mathbb{M}_0^-|-\sqrt{2}P_2(\cos\theta)e^{i\phi_{\text{cc}}^{2-}}|\mathbb{M}_2^-|\right]    
        \\\nonumber
         &- e^{i(\omega\tau+\phi_{\text{XUV}}^+)}  \left[e^{i(\eta_0+\phi_0)}|\mathbb{M}_0^+|+\sqrt{2}P_2(\cos\theta)e^{i(\eta_2+\phi_2)}|\mathbb{M}_2^+|\right].\nonumber
    \end{align}
We define $\chi^\pm(\theta)$ as the argument of the quantities inside the brackets. The angle-resolved RABBIT signal can be written as 
\begin{align}
    I_\text{SB}&(\theta)= |A_\parallel^+(\theta)+A_\parallel^-(\theta)|^2 \propto
    |A_\parallel^+(\theta)|^2+|A_\parallel^-(\theta)|^2    \label{chi}\\ \nonumber &+2|A_\parallel^+(\theta)||A_\parallel^-(\theta)| \cos[2\omega\tau-\Delta \phi_\text{XUV}-\eta_1^--\Delta \chi(\theta)],
\end{align}
where $\Delta \chi(\theta)=\chi^-(\theta)-\chi^+(\theta)$. In contrast to the angle-integrated result, we now have a single oscillatory signal. The price to pay is that the phase depends not only on the energy but also on the emission angle $\theta$. 
Eq.~(\ref{chi}) can be expanded into a Legendre polynomial decomposition as in \cite{Fuchs:20,Peschel}. We will not do that in the present work, but concentrate on the understanding of the phase variation with emission angle.  

\subsection{Angular phase jumps}

\begin{figure}[h]
    \centering
    \includegraphics[width=\linewidth]{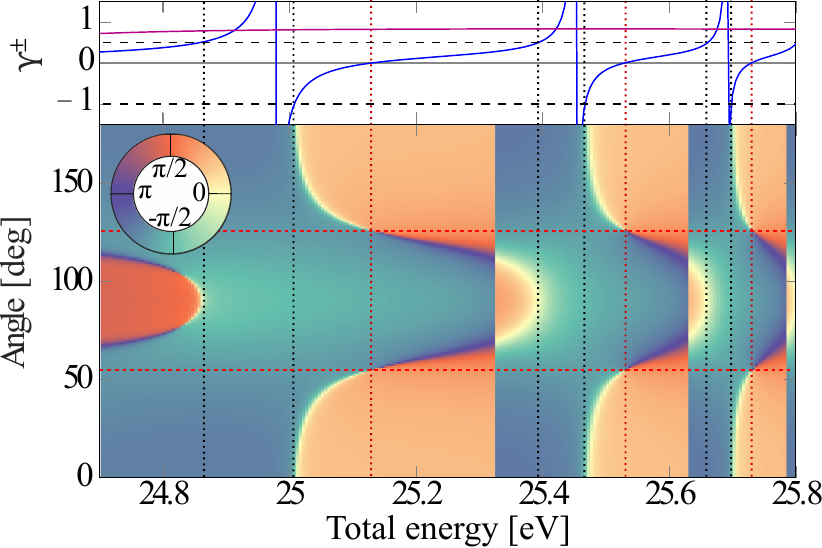}
\caption{Phase variation of the absorption path as a function of angle and energy. The top figure shows the variation of $\Upsilon^\pm$ with energy together with the limits of the inequality in Eq. \ref{ineq} (dashed horizontal black lines). The absorption path is shown in blue and the emission path, which does not lead to any phase jump, in magenta. The black vertical dotted lines indicate the energies at which  $\Upsilon^+$ intersects the black dashed lines. This indicates the limits of the spectral region displaying phase jumps in angle. The vertical red dotted lines indicate the energies at where $\Upsilon^+$ changes sign, leading to a change in the direction of the phase jump in angle. At these energies, the phase jump occurs at the magic angle. The cyclic color bar is shown in the top left corner of the bottom figure. }
    \label{fig:my_label}
\end{figure}

In contrast to what we have discussed so far, which is the phase difference between the absorption and the emission paths [Eqs.~(\ref{chi},\ref{ref})], we now consider the phase difference between the angular channels for the absorption and the emission paths individually,
\begin{align}
&\Delta\varphi^+=\eta_0-\eta_2+\phi_0-\phi_2\\ 
&\Delta\varphi^-=\phi_{cc}^{0-}-\phi_{cc}^{2-}-\pi.    
\end{align}
Factorizing the phase terms, the quantities inside the bracket in Eq.~(\ref{bra}) are proportional to  $\exp(i\Delta \varphi^\pm)|\mathbb{M}_0^\pm|+\sqrt{2}P_2|\mathbb{M}_2^\pm|$. There will be a phase jump in the path considered (emission or absorption) as a function of the emission angle whenever the real part of this expression
changes sign. Introducing
\begin{equation}
    \Upsilon^\pm=\frac{|\mathbb{M}_0^\pm|}{\sqrt{2}|\mathbb{M}_2^\pm|}\cos\Delta\varphi^\pm,
\end{equation}
a phase jump will happen at an angle $\theta$ such that
\begin{equation}
     P_2(\cos\theta)=-\Upsilon^\pm.
\end{equation}
This equation has a solution only if 
\begin{equation}
    - 1\le \Upsilon^\pm \le \frac{1}{2}.
    \label{ineq}
\end{equation}
These two inequalities set conditions for a sign change as a function of angle, and hence a phase jump, in the absorption or emission paths. We also note that when $|\mathbb{M}_0^\pm| \ll \sqrt{2}|\mathbb{M}_2^\pm|$, or when $\Delta\varphi^\pm \approx \pm \pi/2$, a phase jump will happen at an angle such that $P_2(\cos\theta)$ is close to zero, i.e. at a node of $Y_{20}$, $54.7^\circ$ or $125.3^\circ$ (often referred to as ``magic'' angles). In contrast, if $|\mathbb{M}_0^\pm| \gg \sqrt{2}|\mathbb{M}_2^\pm|$ and $\Delta\varphi^\pm \neq \pm \pi/2$, the inequality (\ref{ineq}) is not fulfilled, leading to an absence of a phase jump as a function of angle.
$\Delta\varphi^-$ is positive and, in general, small, increasing close to the threshold \cite{Fuchs:20}. We find that  $\Upsilon^- \ge 1/2$ so that no phase jump can be found in the emission path. 
 The absorption path will in contrast exhibit different energy regions with or without a phase jump, depending on whether Eq.~(\ref{ineq}) is fulfilled or not. Because of the non-zero imaginary part of $\exp(i\Delta \varphi^\pm)|\mathbb{M}_0^\pm|+\sqrt{2}P_2|\mathbb{M}_2^\pm|$, the phase jump as a function of angle is smaller than $\pi$ rad. The direction of the phase jump depends on the sign of $\Delta\varphi^+$. This is illustrated in Figure~\ref{fig:my_label}, showing both $\Upsilon^\pm$ as a function of energy, and the phase variation of the absorption path as a function of angle (vertical axis) and energy (horizontal axis). The black dashed lines indicate the different energy regions, where a phase jump as a function of angle is to be or not be expected.

\subsection{Delay-integrated asymmetry parameters}

The sideband signal can be expressed using the angle-integrated signal, $I_\text{SB}(\tau)$, and two delay-dependent asymmetry parameters ($\beta_i(\tau)$, $i=2,4$) as~\cite{Cooper,busto,Joseph_2020}
\begin{equation}
I_\text{SB}\left(\theta,\tau\right)=\frac{I_\text{SB}(\tau)}{4\pi} 
\left[1+\sum_{i=2,4}\beta_i(\tau)P_i(\cos \theta)\right] 
\label{beta2}
\end{equation}

Another way to parameterize the sideband angular dependence \cite{Saha:2021} is to extract the angle-integrated signal and asymmetry parameters, $I_\text{SB}^\pm$ and $\beta_i^\pm$ for the absorption and emission processes separately, and then similar quantities for the oscillating cross term $A_\parallel^{+*}A_\parallel^-$ (or its complex conjugate).  
The delay-integrated angular $\beta_i$-parameters~\cite{Joseph_2020} can  be simply obtained as \cite{Saha:2021}
\begin{align}
\beta_i = 
\frac{{\beta_i^+ I_\text{SB}^+ +
\beta_i^- I_\text{SB}^-}}
{I_\text{SB}^++I_\text{SB}^-}.
\end{align}
They  are the sum of the ordinary $\beta$-parameters for the absorption and emission paths, weighted with the relative strength of each path. 

\section{Experimental results}

\subsection{Angle-integrated measurements}

\begin{figure}
    \centering
   \includegraphics[width=\columnwidth]{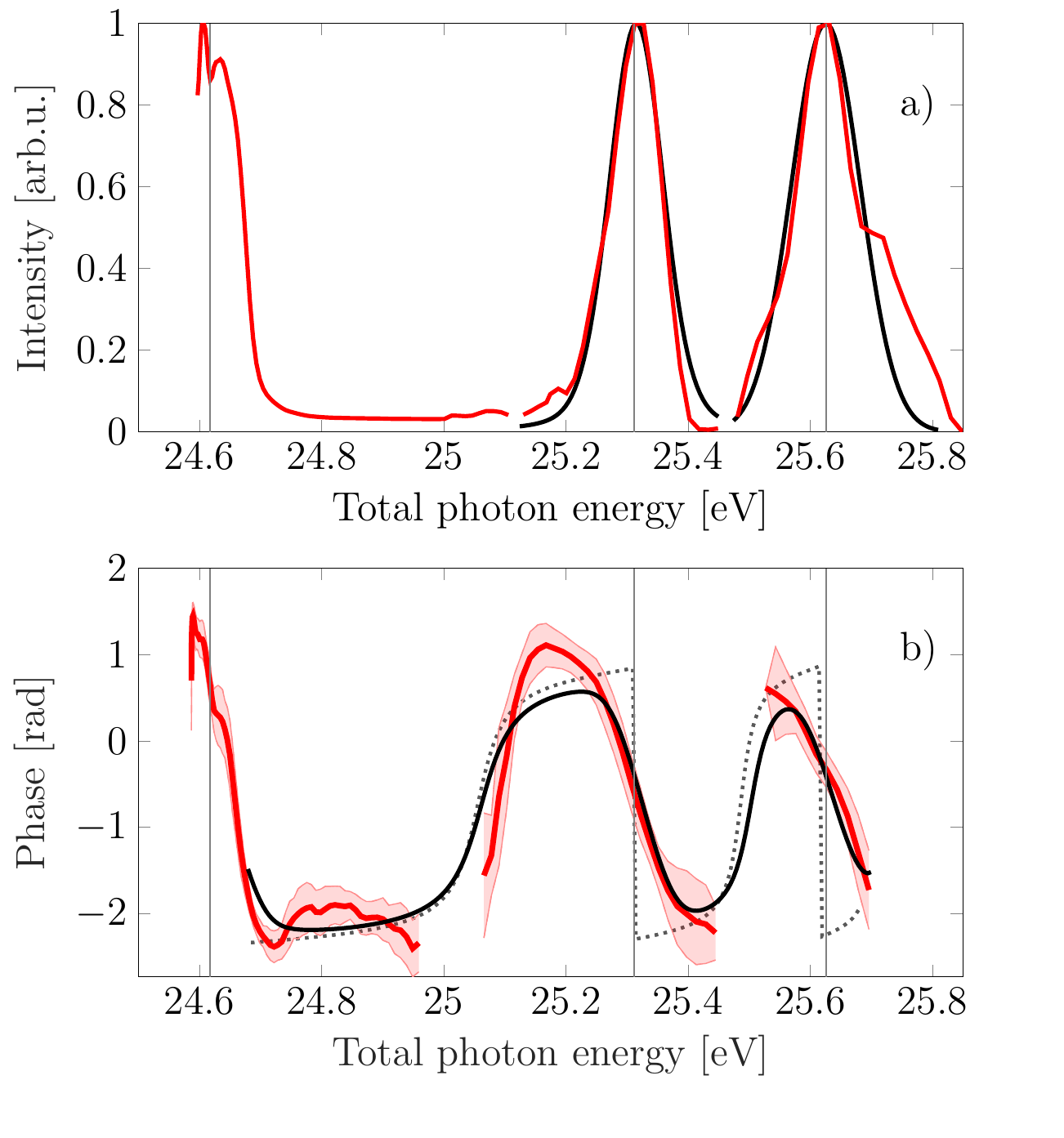}
    \caption{Two-photon ionization via $1snp$ states with $n=3,4,5$. (a) Angle-integrated  intensity. Red line: measurement, black line: RPAE-based simulations. Relative heights of individual peaks are not to scale. (b) Angle-integrated phase. Red solid line: measured data, shaded areas: error of the weighted average. RPAE-based simulations excluding (grey dashed line) and including (black solid line) the effect of the pulse bandwidth and power-induced broadening ($\gamma_p=10$ meV). 
    Vertical grey lines indicate the energies corresponding to the $1snp ^1\text{P}_1$ resonances with $n=3,4,5$.}
    \label{integrated_all}
\end{figure}

Figure \ref{integrated_all} presents results obtained with the velocity map imaging spectrometer. We show the angle-integrated intensity (a) and phase (b) for sideband 16, as a function of the total energy absorbed in the two-photon ionization process. Several measurements have been performed, such that the 15$^\mathrm{th}$ harmonic spans a large energy region across the $ 1snp^1\text{P}_1$ (\textit{n}=3-5) series. 
Figure \ref{integrated_all}(a) presents three broad maxima, approximately centered at the position of the resonances. The relative intensities of the maxima is arbitrary, since the three measurements have been performed with different central laser frequencies. Therefore, we normalize the maxima in Figure \ref{integrated_all}(a). The broadening is partly due to the XUV and IR bandwidths leading to a convolution of the amplitudes (see Eq.~\ref{foldabs}), to the ionization-induced additional width, as well as to the spectrometer resolution. The first peak corresponding to the $1s3p^1\text{P}_1$ resonance exhibits a double structure, that could be due to a continuum structure induced by the interference between direct non-resonant two-photon ionization and ionization from the $1s3p^1\text{P}_1$ bound state \cite{Knight}. Such a feature has been previously observed in photoelectron spectroscopy \cite{autuori}, transient absorption measurements \cite{Chini2013}, as well in numerical simulations \cite{Swoboda}.     
 
Figure \ref{integrated_all}(b) shows the phase variation of the angle-integrated sideband 16 intensity. The presented curves are obtained by performing a weighted averaging over a few measurements, as detailed in the SM. Phase jumps of approximately $-\pi$ rad can be observed when the 15$^\mathrm{th}$ harmonic energy becomes equal to that of the $np^1\text{P}_1$ states. Likewise, phase jumps of opposite sign and similar magnitude can be seen in between each two successive resonant jumps \cite{autuori,Drescher2022}. The splitting of the $1s3p^1\text{P}_1$ state is also accompanied by a phase variation. 

Fig.\;\ref{integrated_all} further shows the results of a two-photon RPAE calculation, not including (dotted black curve, only shown in (b)) and including (black curve) bandwidth effects, with $\sigma_1$= 25 meV, $\sigma_q$= 150 meV, as well as an additional width due to ionization from the Rydberg state ($\gamma_p= 10$ meV). These simulations reproduce the behavior of the intensity and phase variation across the resonances.   

\subsection{Parallel and perpendicular polarizations}

\begin{figure}
    \centering
   \includegraphics[width=0.8\columnwidth]{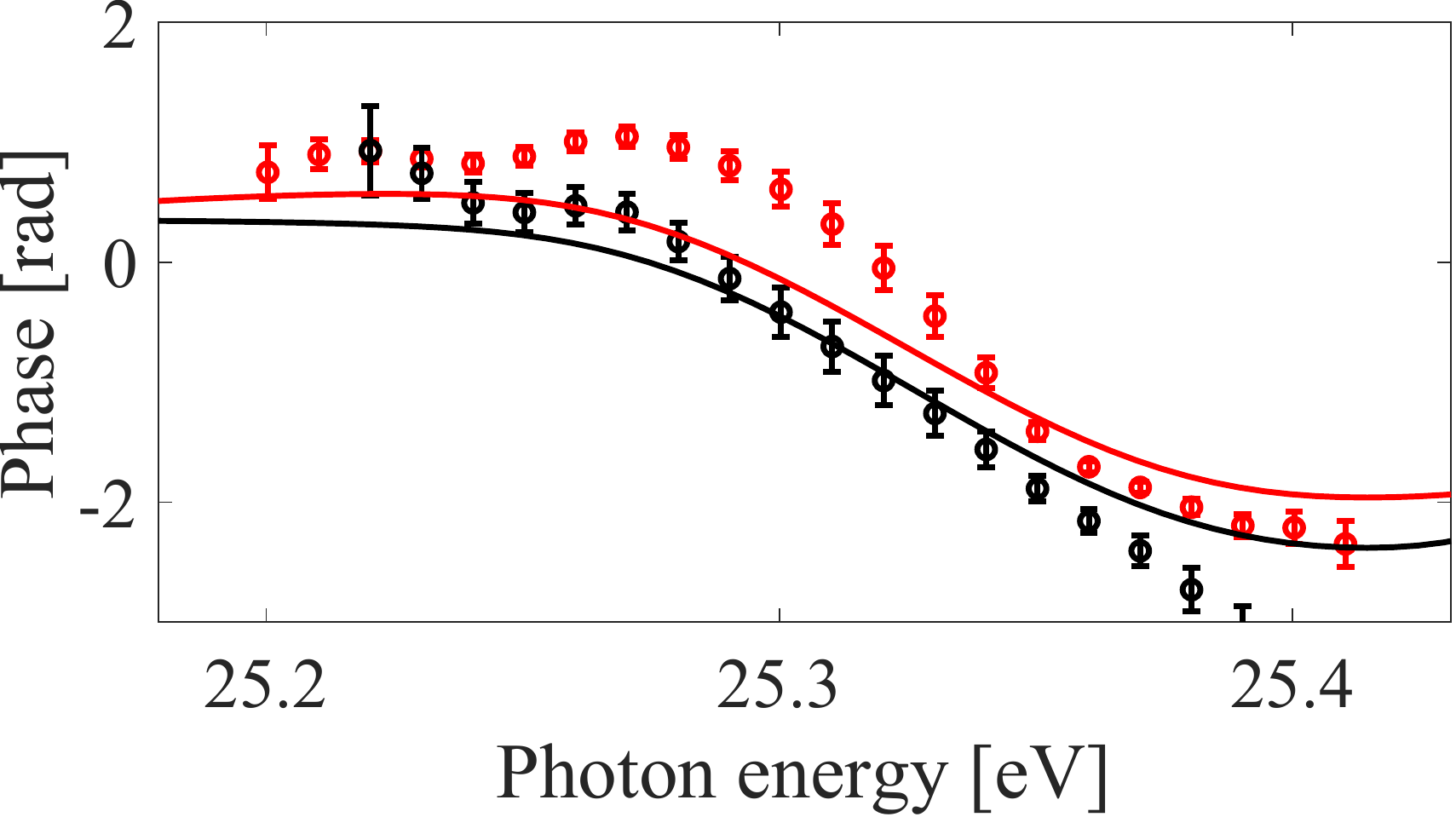}
    \caption{Angle-integrated SB16 phase measured with the magnetic bottle electron spectrometer. Data points: parallel (red) and perpendicular (black) XUV and IR polarizations; solid curves: calculated total phase for parallel XUV and IR polarizations (red) and \textit{d}-wave only phase (black).}
    \label{s_and_d_mbes}
\end{figure}

High-spectral-resolution angle-integrated RABBIT measurements were performed in the energy region around the $1s4p^1\text{P}_1$ resonance.
Our aim was to compare measurements with the dressing field polarization parallel ($0^\circ$) and perpendicular ($90^\circ$) to that of the XUV, while keeping the probe intensity at the same value. Both \textit{s} and \textit{d} channels coexist in the parallel configuration, while only the \textit{d} channel is allowed when the probe polarization is perpendicular to that of the pump, since only $\Delta m = \pm1$ transitions are possible. In these measurements, we normalize the phase variation to that of the non-resonant sideband 18, which exhibits a flat phase. A comparison between parallel and perpendicular IR polarizations is shown in Figure~\ref{s_and_d_mbes}. The black and red symbols and solid curves correspond to the parallel and perpendicular case, respectively. We observe a smooth phase variation across the resonance rather than a sudden $\pi$ jump, in agreement with the results shown in Figure \ref{angle-int}. In order to reproduce this behavior in theoretical calculations, it is necessary to include both ionization broadening and folding due to finite pulse effects. In the region investigated, close to the $1s4p$ resonance, the phase variation of the $s$ and $d$ channels are very similar, so that the parallel or perpendicular configurations do not yield very different phase measurements.

\subsection{Delay-integrated photolectron angular distributions}

\begin{figure}
\centering
   \includegraphics[width=0.8\columnwidth]{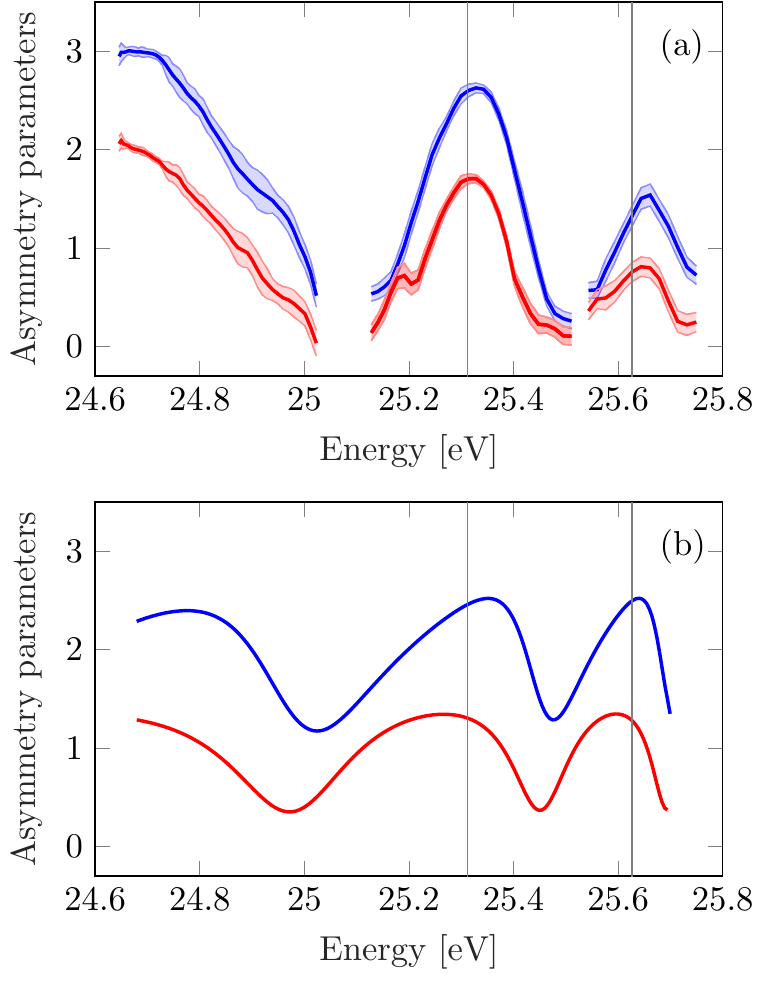}
    \caption{Energy variation of the delay-integrated asymmetry parameters $\beta_2$ (blue) and $\beta_4$ (red); (a) Experimental results; (b) Theoretical calculations obtained at a given dressing field wavelength, equal to 782 nm. The grey lines indicate the energies corresponding to the $1snp$ $^1$P$_1$ resonances with n=4,5.}
    \label{beta}
\end{figure}

Figure \ref{beta}(a,b) presents the variation of the $\beta_2$ and $\beta_4$ parameters for the delay-integrated signal. These coefficients vary significantly across the observed energy range, with maxima close to the $1s3p^1\text{P}_1$, $1s4p^1\text{P}_1$ and $1s5p^1\text{P}_1$ resonances. The experimental behavior is reproduced by the calculations, here shown without including any folding, i.e. assuming a monochromatic dressing field, at a wavelength of 782 nm. The strong variation of the $\beta_2$ and $\beta_4$ parameters between two resonances indicates a varying angular distribution, due to a change in the relative amplitude and phase of the $s$ and $d$ angular channels. 

\subsection{Angle-resolved phase measurements}

\begin{figure}
    \centering
   \includegraphics[width=\columnwidth]{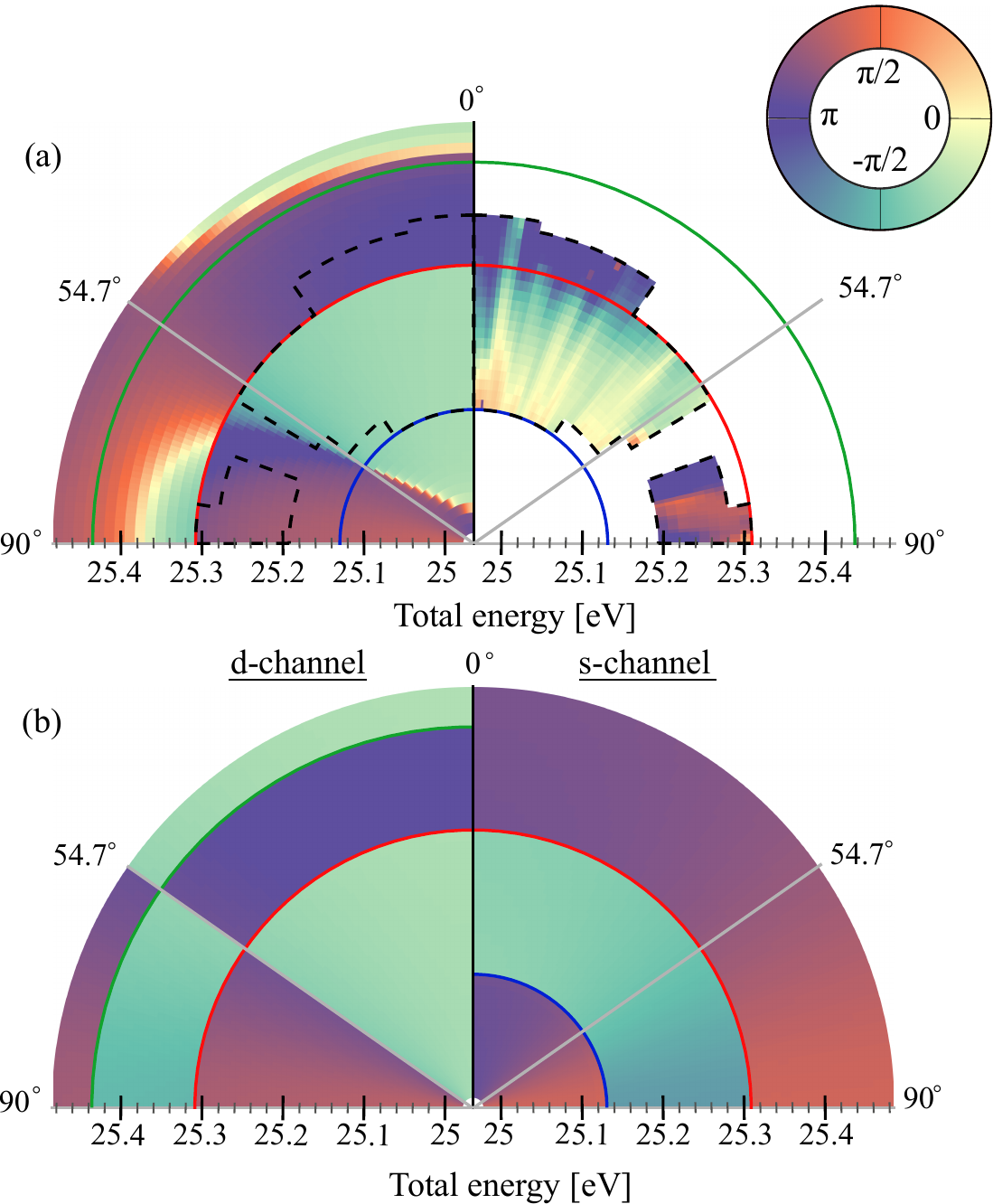}
    \caption{ Angle-and-energy resolved relative phase of SB16 around the 4p resonance. (a) Comparison between experiment (right) and theory (left). Only the experimental results with significant signal-over-noise ratio are presented. This region is indicated by the black dashed line. The radial coordinate is the total energy, while the polar angle is the electron emission angle relative to the polarization axis. The cyclic color bar is shown in the top right of the figure. In (b), we present the contribution of the $s$ (right) and $d$ (left) waves in the absorption path to the total phase variation.}
    \label{4p}
\end{figure}

We here concentrate on the spectral region around the $1s4p^1\text{P}_1$ resonance. Figure~\ref{4p}  displays the phase variation as a function of angle and energy. The phase is indicated in color, following a cyclic representation, thus avoiding unwrapping issues. The phase decreases (increases) when moving clockwise (anti-clockwise) in the cyclic color bar. Fig.\;\ref{4p}(a,right) shows the measured values in the angle and energy region, delimited by a dashed contour, where the signal is sufficiently high for a reliable phase and amplitude retrieval. Fig.\;\ref{4p}(a,left) shows the results of the two-photon-RPAE calculation. The two results agree quite well, and in the following, we focus on the interpretation of the theoretical results, which give a more complete picture of the phase variation as a function of both energy and angle. In order to gain a deeper insight into the interplay of the different angular channels, we show the sideband variation where we selectively suppress the resonant path to the $s$ continuum (b,left) or to the $d$ continuum (b,right).

In general, the phase variation observed in Fig.~9(a,left) is very similar to that in Fig.~9(b,left), indicating that the overall phase variation in angle and energy is dominated by that of the resonant path to the $d$ wave. In the following we focus on interpreting the phase variation in the vicinity of three spectral regions indicated with coloured half circles in Fig.~9(a).

First at the position of the 4$p$ resonance, indicated by a red half circle close to 25.31 eV, a sharp $\pi$ rad phase jump along the energy axis is observed at all emission angles. In this region both the $s$ and $d$ channels experience a $\pi$ jump due to the resonance such that angle resolved measurements do not provide additional information compared to angle integrated measurements.

Another phase jump along the energy axis can observed in the vicinity of the green half circle. In this region, the matrix element $\mathbb{M}_2^+$ crosses 0 and changes sign such that the relative weight of the $s$ and $d$ resonant paths varies very fast as a function of energy. As a result, a strong angular dependence of the position of the phase jump can be observed in Fig~9(a,left): at angle above 54.7 degrees the phase jump appears above the green line, while at angles below the magic angle, the phase jump appears at energies below the green line. This is consistent with the fact that $\mathbb{M}_2^+$ and $Y_{20}$, should have opposite signs in order for the real part of $A^+_\parallel(0)$ to cancel.

 At 54.7 degrees, where $Y_{20}$ has a node, the phase variation as a function of energy is fully dominated by the $s$ channel as can be observed by comparing Fig. 9(a,left) with Fig. 9(b,right). In addition, in the close vicinity of the spectral region where $\mathbb{M}_2^+$ cancels no phase variation as a function of emission angle is observed in (a,left), since the $s$ channel dominates. This corresponds to the spectral region shown in Fig. 5, where the inequalities in Eq.~\ref{ineq} are not fulfilled. 
 
In contrast to the region close to the 4$p$ resonance, here angle resolved measurements provide a lot of information on the competition between the $s$ and $d$ resonant channels that cannot be obtained from angle-integrated measurements. 

Finally, over a large energy range, from 25~eV to 25.3~eV, a sharp phase jump can be observed as a function of angle close to 54.7 degrees. In this spectral region the $d$ resonant channel strongly dominates and the phase jump as a function of angle reflects the change of sign of $Y_{20}$. The exact angle at which the phase jumps and the magnitude of the phase jump depends on the relative weight between $s$ and $d$. When $\mathbb{M}_0^+=0$ (blue circle), the phase exhibits a sharp pi jump exactly at the magic angle, while the phase jumps occurs at lower(higher) angles below(above) the blue line. Interestingly, although the d channel dominates, the change of sign of $\mathbb{M}_0^+$ across the blue line manifests as a change of direction of the angular phase jump.

This analysis shows that by examining the positions of the phase jumps as a function of angle and energy, it is possible to infer not only the position of the resonance, but also the positions of the sign changes of $\mathbb{M}_L^+$, even in the case where two-photon ionization is very much dominated by one contribution, the $d$ channel. This also shows that phase measurements as a function of angle and energy give a lot of information about the relative contributions of the two angular channels.    

\section{Conclusion} 

In conclusion, we have studied the phase and intensity of two-photon ionization through the  $1s3p^1\text{P}_1$, $1s4p^1\text{P}_1$ and $1s5p^1\text{P}_1$ resonance series, using an interferometric method. While the intensity variation presents a series of peaks at the positions of the resonance, the phase undergoes $\pi$-rad phase changes both at and in between the resonances. Going through the resonance, the phase decreases linearly with energy with a total phase variation of $\pi$ rad at all emission angles. The smaller the ionization induced by the dressing field, the more rapid this linear decrease in phase is. The phase variation remains similar when varying the relative polarization of the XUV and dressing fields. A second sharp phase variation of $\pi$ rad is observed between two resonances. This phase variation is strongly dependent on the emission angle and reflects the angle- and polarization- dependent interplay between the two angular channels contributing to the two-photon ionization process.        

We have shown both experimentally and theoretically that, even for a simple atomic system like helium, angle-resolved measurements, as well as measurements with different polarizations, provide invaluable insight to the physics of resonant two-photon ionization. They should become an essential tool to the investigation of more complex atoms or molecules. 

\section*{Acknowledgements}

The authors acknowledge support from the Swedish Research Council (2013-8185, 2016-04907, 2017-04106, 2018-03731, 2020-0520, 2020-03315, 2020-06384), the Swedish Foundation for Strategic Research (FFL12-0101), the European Research Council (advanced grant QPAP, 884900) and the Knut and Alice Wallenberg Foundation. AL and MA are partly supported  by the Wallenberg Center for Quantum Technology (WACQT) funded by the Knut and Alice Wallenberg foundation. JL acknowledges financial support by the European Union’s Horizon 2020 research and innovation program under the Marie Skłodowska-Curie Grant Agreement 641789 (MEDEA).

% \bibliography{ms.bbl}%,Ref_lib.bib}


\end{document}